\documentclass[twocolumn,preprintnumbers,superscriptaddress,landscape,nofootinbib]{revtex4}
\usepackage[utf8]{inputenc}
\usepackage{amsmath}
\usepackage{amssymb}
\usepackage{amsthm}
\usepackage{float}
\usepackage{braket}
\usepackage{graphicx}
\usepackage{physics}
\usepackage{url}
\usepackage[colorlinks=true, pdfstartview=FitV, linkcolor=blue, citecolor=blue, urlcolor=blue]{hyperref}
\usepackage{slashed}
\usepackage[english]{babel}
\usepackage[normalem]{ulem}
\usepackage{tikz}
\usepackage{quantikz}
\graphicspath{{./figures/}}
\usepackage{fdsymbol}
\usepackage{cases}

\begin{document}

\title{Quantum Protocol for Decision Making and Verifying Truthfulness among $N$-quantum Parties: Solution and Extension of the Quantum Coin Flipping Game}
\author{Kazuki Ikeda}
\email[]{kazuki7131@gmail.com}
\affiliation{Co-design Center for Quantum Advantage (C2QA), Stony Brook University, USA}
\affiliation{Center For Nuclear Theory, Department of Physics and Astronomy, Stony Brook University, USA}

\author{Adam Lowe}
\email[]{a.lowe3@aston.ac.uk}
\affiliation{College of Engineering and Physical Sciences, School of Engineering and Applied Science, Aston University, Birmingham B4 7ET, United Kingdom}



\begin{abstract}
We devised a protocol that allows two parties, who may malfunction or intentionally convey incorrect information in communication through a quantum channel, to verify each other's measurements and agree on each other's results. This has particular relevance in a modified version of the quantum coin flipping game where the possibility of the players cheating is now removed. Furthermore, the analysis is extended to $N$-parties communicating with each other, where we propose multiple solutions for the verification of each player's measurement. The results in the $N$-party scenario could have particular relevance for the implementation of future quantum networks, where verification of quantum information is a necessity.
\end{abstract}
\maketitle

\section{Introduction}
\subsection{General Background}
Quantum communication using quantum channels is becoming practical, but there are many issues that need to be addressed in order to actually operate them in business. In the usual setting of multiparty secure computation, many protocols assume secure communication channels between all two parties~\cite{goldreich2019play,ben2019completeness}. For example, the conventional quantum key distribution~\cite{2020arXiv200306557B} realizes secure keys for secret channels under the assumption that the sender and the receiver are trusted. However, it is non-trivial for a player to achieve a reliable communication channel without trusting other parties.

Besides, regardless of whether one can trust the other party or not, even if secure communication channels are realized, all kinds of problems can occur in real human communication, including political and business problems. While it is important to pursue quantum channel technology that allows for secure and accurate quantum communication, it is equally important to pursue the design of software and systems that will operate correctly and desirably on the quantum channel.

The study of desirable systems and software for people on a network/market is called mechanism design (or market design) and is widely studied in economics in terms of auctions, optimal matching, and allocation of public goods~\cite{milgrom2004putting,10.2307/1911013,10.2307/1817047,10.1111/1467-937X.00076,10.2307/3689266}. (The Sveriges Riksbank Prize in Economic Sciences in Memory of Alfred Nobel 2007 was awarded to Hurwicz, Maskin and Myerson
“for having laid the foundations of mechanism design theory”~\cite{Nobel2007}.) 

In this study, we consider the quantum coin flipping game~\cite{2020arXiv200306557B} from a perspective of mechanism design. This paper assumes that quantum channels are in practical use. We discuss the problems that players who play the quantum coin flipping game may face, and redesign the game. When looking at the quantum coin flipping game from this viewpoint, it is not so much a matter of cryptography, but rather a matter of consensus building, which refers to the process essential for building consensus among the parties involved in the same assignment, project, or business meeting with a customer. Consensus building is essential for sharing objectives and facilitating business.

Extending a two-person quantum coin flipping game to an $N$-person game is indeed the General Byzantine Problem, which asks whether a group of communicating objects as a whole can form a correct consensus when communication or individual objects may convey incorrect information due to malfunction or intentionality. This issue has been studied in a wide range of fields~\cite{10.1145/357172.357176,10.1145/322186.322188}, including Blockchain~\cite{bitcoin,ZHANG20181,BANERJEE201869,IKEDA201899} and its quantum extensions~\cite{IKEDA2018199,10.1007/978-3-030-01174-1_58}. The $N$-party coin flipping game that we presented is a simple case of a generic Byzantine Generals Problem or the $N$-party decision making problem. Our quantum coin flipping problem for $N$ people is exactly the quantum Byzantine General Problem when each quantum general has a single qubit.

To design a quantum game that can work properly as a game, let us consider the conditions that a game must meet. In designing a game, the minimum requirements to be met would include
\begin{itemize}
\item[\textcolor{blue}{(A)}] There is no room for cheating.
\item[\textcolor{blue}{(B)}] Each player is correctly aware of the other's results.
\item[\textcolor{blue}{(C)}] Each player can agree on the outcome of the game.
\end{itemize}
Condition \textcolor{blue}{(A)} is necessary to achieve consistency in the rules of a game. If there is a loophole in the rules, an attack that exploits a vulnerability in the rules is possible. There is no reason to believe that the remote players who are about to play against each other will not cheat. The use of device-independent quantum key delivery, which is effective even when the sender and receiver are not trusted, could solve this problem~\cite{mayers1998quantum,10.5555/2011827.2011830,acin2007device,PhysRevLett.128.110506}. However, that is not the only problem. It is even more difficult to achieve a system where everyone can agree on the outcome, even if the game is played correctly. 

Condition \textcolor{blue}{(B)} is necessary for each player to confirm the progress of the game. This allows one to verify the validity of one's past strategies and to plan for future strategies. Clearly, having the correct information about the other players is important when building consensus.

Condition \textcolor{blue}{(C)} is necessary for the game to converge. At the end of the game, the outcome is determined and all players accept the result. Players would lose incentive to participate in a game where the outcome of the game is controversial. Besides, even if the players temporarily agree on the outcome, they may later reverse their agreement. In such cases, it is necessary in practice to make it provable to a third party that an agreement has been reached in order to avoid a situation of incomplete contracts~\cite{coase,10.2307/725234,RePEc:ecm:emetrp:v:52:y:1984:i:2:p:449-60,10.2307/1912698,RePEc:ucp:jpolec:v:94:y:1986:i:4:p:691-719}.

The link between game theory and quantum mechanics was founded over 20 years ago, and the development in the field since its initial discovery has been rapid ~\cite{1999PhRvL..83.3077E, PhysRevLett.82.1052,Iqbal2000EvolutionarilySS,Piotrowski2001QuantumMG,PhysRevA.65.022306}. Alongside this, there has been research into quantum information with specific focus on quantum networks and its potential implementation in quantum computing~\cite{kimble2008quantum,Elliott_2002,10.1117/12.606489,Peev_2009,Sasaki:11,dynes2019cambridge}. In recent years, these two fields have started to work in unison where the advantages gained from quantum game theory can be utilised for the benefit of quantum networks. 

Quantum games have been developed for repeated games~\cite{2020QuIP...19...25I,2021QuIP...20..387I}, extensive form games~\cite{2022arXiv220705435I}, contract theory~\cite{ikeda2021quantum}
and markets in quantum networks~\cite{ikeda2022theory}. This has a natural crossover with quantum mechanics due to uncertainty being prevalent in both fields. From this, it is clear that the potential advantage that can be gained from using quantum correlations in network systems could have significant practical implications in quantum technologies.

There are a wide range of quantum games that have been investigated where quantum correlations have been found to yield quantum advantage compared to the respective classical counterpart. A common example for this is the CHSH game, which allows a practical implementation for the benefit of non-locality. Interestingly, it was found that there is an inherent link between Bell's inequalities and Bayesian game theory~\cite{bellbayes}.

The game which this paper focuses on is the quantum coin flipping game \cite{2020arXiv200306557B}, where the essence of the game is based on a two-player, two-outcome game.

\subsection{Statement of Results}
In this work we formulated the quantum coin flipping game with two parties and extended it to a game with $N$ parties. We have ensured that any possibility of cheating or attack is eliminated, under the assumption that each player only flips a coin and does not operate arbitrarily on their own quantum state. 

It is important to recall that in the conventional quantum coin flipping game, even with these natural assumptions imposed, there was still plenty of room for remote players to cheat (Problems \textcolor{red}{($\varheartsuit,\clubsuit,\spadesuit$)} defined later). In the conventional game, regardless of the outcome of the coin flip, the player who announces the result last always can cheat and therefore always can win, and this can be regarded as an ultimatum game. To avoid the quantum coin flipping game becoming an ultimatum game, we re-designed the game using an entangled state~\eqref{eq:initial} between the players (Fig.~\ref{fig:circuit}). As we described in the main text, we solved those three problems~\textcolor{red}{($\varheartsuit,\clubsuit,\spadesuit$)}. 

\section{Quantum coin flipping}
\subsection{Preliminaries}
The quantum coin flipping game originates from the classical coin flipping game where there are two parties, which will be denoted by Alice ($A$) and Bob ($B$). Consider the scenario where Alice and Bob are a recently divorced couple who decide to play a game to determine who gets the car they previously shared. Since they do not like each other, they are living far away from each other, so decide to play this game over the telephone. The game is set up as follows: both players have a fair two-sided coin which can either land on heads or tails. They then each flip their respective coins. If one player lands on heads, and the other lands on tails, the player who landed on heads wins. If they both land on heads, they flip again, and if they both land on tails, they flip again, until there is a winner. However, it is clear that each player cannot verify the other players result, therefore when communicating over the telephone, if Alice claims to have landed on tails, then Bob can win the game by claiming to have landed on heads, even if this were not the case. This game can be implemented as shown in Fig.~\ref{fig:Classical}, whose sketch is shown in Fig.~\ref{fig1}. As can be easily seen from the figures, when such a game is played without sharing information with each other, the last person to claim the outcome can always become the winner. Such games are called ultimatum games~\cite{doi:10.1177/002200276100500205}.
\begin{figure}[H]
    \centering
\begin{quantikz}
\ket{0}_A  &\gate{H}\gategroup[2,steps=1,style={dashed,rounded corners,fill=blue!10,inner xsep=2pt},background]{{Prepare coins}} &\qw&\qw& \meter{}\gategroup[2,steps=1,style={dashed,rounded corners,fill=yellow!30,inner xsep=2pt},background]{{ Flip coins}}\\
\ket{0}_B&\gate{H}&\qw&\qw& \meter{}
\end{quantikz}
    \caption{Conventional Setting of the traditional Coin Flipping Game.}
    \label{fig:Classical}
\end{figure}
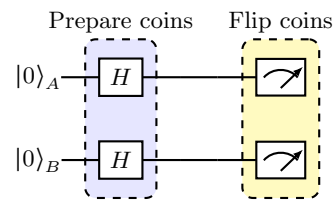
\begin{figure}[H]
    \centering
    \includegraphics[scale=0.35]{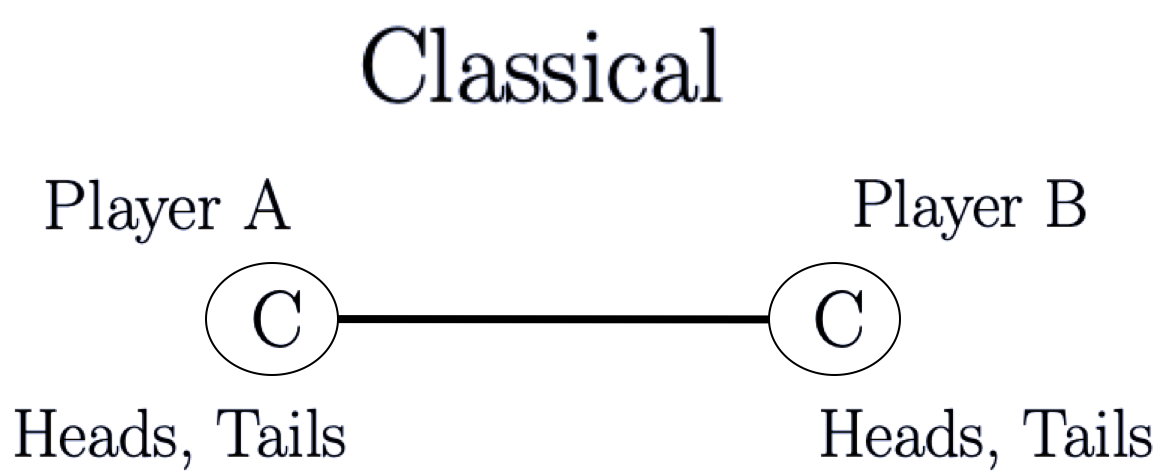}
    \caption{The classical coin flipping game can be represented by the figure above. Both players flip their respective coins, and based on their measurements, they then communicate their results to each other. From this the players decide on the winner of the game. Therefore, it is clear that both players must trust each other to report their results correctly, otherwise the game could be unfairly manipulated.}
    \label{fig1}
\end{figure}

The quantum version of this game is similar, but in this scenario Alice prepares a quantum state in addition to their coin. Whether Alice lands on heads or tails determines what basis Alice measures their quantum state in, and from this, Alice sends Bob this prepared quantum state. Bob then performs measurements on this quantum state, and attempts to deduce what basis Alice measured in. Bob then sends this state back to Alice in addition to communicating to Alice what basis Bob believes the state was prepared in, and from this, Alice can confirm whether Bob deduced the measurement basis correctly, and thus whether the coin landed heads or tails. This can work in the opposite direction, where Bob prepares the quantum state and measures in a particular basis and sends this quantum state to Alice. Due to the quantum correlations, this form of the coin flipping game does allow Alice and Bob to increase their chances of winning. However, this game still allows the possibility of cheating, as ultimately Alice and Bob still have to perform classical communication along the channel.

\subsection{What were the problems?}
In this section we present our solution for two-party quantum coin flipping game. To describe our contribution accurately, let us elaborate on what the challenges were and how they were solved. The remaining challenges of the previous research on this issue has been to establish a way to fairly and rigorously recognize each other's independent results between two remote parties. From the perspective of a mechanism design, this game setting has the following problems.
\begin{enumerate}
    \item[\textcolor{red}{($\varheartsuit$)}]The two people can chose arbitrary coins independently.
    \item[\textcolor{red}{($\clubsuit$)}] The result of one cannot be recognized by the other.
    \item[\textcolor{red}{($\spadesuit$)}] There is a time lag between when one player flips a coin and when the other player learns the result.
\end{enumerate}
Under these circumstances, it is obvious that they could cheat in any way they want. Despite the simplicity of the problem set-up, the second and the third reasons make this problem difficult and fairness between players is lost.

The first point~\textcolor{red}{($\varheartsuit$)} concerns fairness before the game is played. The presence or absence of prior information about the tools used in the game can lead to information asymmetry, which is an important factor in the progression of the game. 
When information about the game is asymmetric, the benefits of changing the rules of the game vary from player to player. A game in which rule changes are impossible is not a good game because it lacks flexibility and development. For each player to be allowed to independently select any coin, knowledge of which coin is selected must be disclosed to all players. However, this is not possible in a setting where each player is remote and there is no neutral third party to verify this.

In the quantum coin flipping game, the second problem~\textcolor{red}{($\clubsuit$)} is even more serious than the first. This problem occurs immediately after the game begins and before communication with others begins. In the conventional setting of quantum coin flipping, it is possible to lie to the other player since only the player can observe their own result. This makes the game no longer work because each player can arbitrarily change the outcome, eliminating the need for a coin toss in the first place. In other words, it is no longer even a "coin-flipping game", as there is no need to even prepare coins in the first place. 

The third problem~\textcolor{red}{($\spadesuit$)} exacerbates the second. In the real physical environment, the speed at which information is transmitted to the other party is finite, so it takes a finite amount of time to convey the result of a coin to a remote party. The presence of this time difference would be enough to cause hesitation and frustration to the players. Even if players decide to play fair at the start of the game, they may decide to take advantage of the time difference to change their results after their own results are observed. Moreover even if all players were honest about the results, the existence of the time difference is sufficient to lead one to believe that changes were made to the results. Mail-in ballots in presidential elections, for example, contribute to creating this kind of distrust among some people.

\section{Solution for two-party quantum coin flipping}
\subsection{How we solved the problems}
As we have discussed, the conventional quantum coin flipping game is fundamentally flawed in its setup, which not only prevents the game from being executed properly, but also makes it unworkable from the start. 

Therefore, we first need to redesign the game so that it can be played correctly. To this end, let us first look back at what the fundamental concept of the game was:
\begin{enumerate}
    \item Each of the two remote players flips a coin.
    \item A winner is determined based on the results of the coins observed by each player.  
\end{enumerate}

Now let us design the quantum coin flipping game as follows. In order to remove the possibility of cheating, consider now that Alice and Bob perform their measurements on a shared entangled state given by
\begin{equation}
\label{eq:initial}
    \ket{\psi}_\text{Coin} = \sum_{ij} c_{ij} \overbrace{\underbrace{\textcolor{blue}{\ket{i}_A}}_{\text{A's coin}}\textcolor{blue}{\otimes} \underbrace{\textcolor{blue}{\ket{j}_B}}_{\text{B's coin}}}^{\text{Flipping}}\otimes\overbrace{\underbrace{\textcolor{red}{\ket{j}_A}}_{\text{B's result}}\textcolor{red}{\otimes}\underbrace{\textcolor{red}{\ket{i}_B}}_{\text{A's result}}}^{\text{Confirmation}},
\end{equation}
where $\sum_{ij}|c_{ij}|^2=1$. The first two qubits $\textcolor{blue}{\ket{i}_A\otimes\ket{j}_B}$ correspond to coins of Player $A$ and Player $B$. For example, Player $A$ can flip a coin by observing their own qubit $\textcolor{blue}{\ket{i}_A}$. The remaining two qubits $\textcolor{red}{\ket{j}_A\otimes\ket{i}_B}$ record the results of Player $A$ and $B$, respectively.
For example, Player $A$ can confirm Player $B$'s result by observing their own qubit $\textcolor{red}{\ket{j}_A}$.

At this point, the probability that Player $A$ observes $\textcolor{blue}{\ket{i}_A}$ is 
\begin{equation}
\label{eq:PlayerA}
\text{Prob}_A(i)=\sum_j|c_{ij}|^2.
\end{equation}
Similarly the probability that Player $B$ observes $\textcolor{blue}{\ket{i}_B}$ is
\begin{equation}
\label{eq:PlayerB}
\text{Prob}_B(i)=\sum_j|c_{ji}|^2.
\end{equation}
The game consists of 4 stages and proceeds as follows.
\begin{description}
    \item[1. Preparation of an entangled state:] Prepare an initial entangled state $\ket{\psi}_\text{Coin}$~\eqref{eq:initial}.
    \item[2. Coin flipping stage:] Each player independently makes a measurement on their own coin qubit $\textcolor{blue}{\ket{i}_A\otimes\ket{j}_B}$. 
    \item[3. Confirmation stage:] Each player independently confirms the opponent's state by measuring their own second (ancilla) qubit $\textcolor{red}{\ket{j}_A\otimes\ket{i}_B}$.
    \item[4. Decision making stage:] Each player compares the results of their own coins with those of their opponents to recognize and agree on the winners and losers.
\end{description}

Here we explain how this protocol works.
Let us first make sure that this game does not depend on the order in which the players play. Of course, they can play simultaneously. Suppose Player $A$ flips a coin and gets $\textcolor{blue}{\ket{i}_A}$. Then the state of the coin~\eqref{eq:initial} changes into  
\begin{equation}
\label{eq:flip}
\ket{\psi}_\text{Coin} \to \sum_{ij}c_{ij}\overbrace{\underbrace{\textcolor{blue}{\ket{j}_B}}_{\text{B's coin}}}^{\text{Flipping}}\otimes\overbrace{\underbrace{\textcolor{red}{\ket{j}_A}}_{\text{B's result}}\textcolor{red}{\otimes}\underbrace{\textcolor{red}{\ket{i}_B}}_{\text{A's result}}}^{\text{Confirmation}}.
\end{equation}
Now let Player $B$ flip a coin. It is easy to see that Player $A$'s result does not affect the probability distribution~\eqref{eq:PlayerB} of Player $B$. If Player $B$'s coin is $\textcolor{blue}{\ket{j}_B}$, the state of the game changes from \eqref{eq:flip} into
\begin{equation}
\ket{\psi}_\text{Coin}\twoheadrightarrow\overbrace{\underbrace{\textcolor{red}{\ket{j}_A}}_{\text{B's result}}\textcolor{red}{\otimes}\underbrace{\textcolor{red}{\ket{i}_B}}_{\text{A's result}}}^{\text{Confirmation}}.
\end{equation}

In \textbf{Confirmation stage}, each player can confirm the state of their opponent by measuring the corresponding ancilla state. For example, Player $A$ finds $\textcolor{red}{\ket{j}_A}$ for the result of Player $B$ with probability 1 and vice versa. 

\begin{figure}[H]
    \centering
\begin{quantikz}
\ket{0}_A   &\gate{H}\gategroup[4,steps=2,style={dashed,rounded corners,fill=blue!10,inner xsep=2pt},background]{{Prepare coins}}& \ctrl{1} &\qw & \meter{}\gategroup[3,steps=1,style={dashed,rounded corners,fill=yellow!30,inner xsep=2pt},background]{{ Flip coins}}\\
\ket{0}_A&\qw& \targ{} & \swap{2}\gategroup[3,steps=1,style={dashed,rounded corners,fill=green!30,inner xsep=2pt},background, label style={label position=below,anchor=north,yshift=-0.2cm}]{{\sc swap}}& \qw&\meter{}\gategroup[3,steps=1,style={dashed,rounded corners,fill=red!20,inner xsep=2pt},background, label style={label position=below,anchor=north,yshift=-0.2cm}]{{Confirm results}}\\
\ket{0}_B &\gate{H}& \ctrl{1} &\qw &\meter{}\\
\ket{0}_B&\qw& \targ{} &\targX{} & \qw&\meter{}
\end{quantikz}
    \caption{Quantum Circuit for Quantum Coin Flipping Game.}
    \label{fig:circuit}
\end{figure}
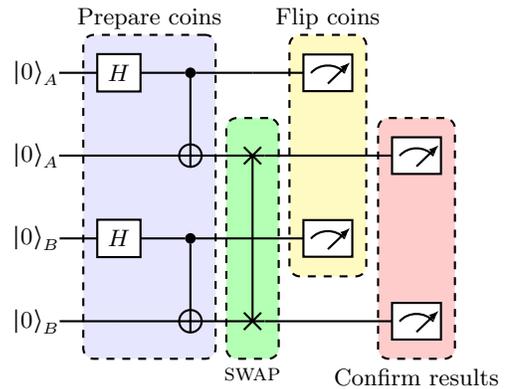

For simplicity, let us play the classical setting. Then the state \eqref{eq:initial}, which we use for the game is generated by the circuit shown in Fig.\ref{fig:circuit}, in which coins are prepared by the following procedure
\begin{align}
\begin{aligned}
\ket{\textcolor{blue}{0}\textcolor{red}{0}\textcolor{blue}{0}\textcolor{red}{0}} \to& \frac{1}{2}\big(\ket{\textcolor{blue}{0}\textcolor{red}{0}}_A\ket{\textcolor{blue}{0}\textcolor{red}{0}}_B+\ket{\textcolor{blue}{0}\textcolor{red}{0}}_A\ket{\textcolor{blue}{1}\textcolor{red}{1}}_B \\ &+\ket{\textcolor{blue}{1}\textcolor{red}{1}}_A\ket{\textcolor{blue}{0}\textcolor{red}{0}}_B +\ket{\textcolor{blue}{1}\textcolor{red}{1}}_A\ket{\textcolor{blue}{1}\textcolor{red}{1}}_B\big)\\
\to&\frac{1}{2} \big(\ket{\textcolor{blue}{0}\textcolor{red}{0}}_A\ket{\textcolor{blue}{0}\textcolor{red}{0}}_B+\ket{\textcolor{blue}{0}\textcolor{red}{1}}_A\ket{\textcolor{blue}{1}\textcolor{red}{0}}_B \\ &+\ket{\textcolor{blue}{1}\textcolor{red}{0}}_A\ket{\textcolor{blue}{0}\textcolor{red}{1}}_B+\ket{\textcolor{blue}{1}\textcolor{red}{1}}_A\ket{\textcolor{blue}{1}\textcolor{red}{1}}_B \big),
\end{aligned}
\end{align}
where we replaced $\ket{0}\leftrightarrow\ket{\uparrow}$ and $\ket{1}\leftrightarrow\ket{\downarrow}$.

Player $A$'s \textcolor{blue}{first qubit} is a coin of Player $A$, who can use the \textcolor{red}{second qubit} to confirm Player $B$'s result.

\subsection{Validity of the Protocol}
Here we show the validity of our protocol and confirm that the problems \textcolor{red}{($\varheartsuit$)}, \textcolor{red}{($\clubsuit$)} and \textcolor{red}{($\spadesuit$)} raised in the previous section are completely solved in principle. 

Regarding the problem \textcolor{red}{($\varheartsuit$)}, they both use the same state \eqref{eq:initial} to play the game. The probability distribution is completely determined by the matrix 
\begin{equation}
\label{eq:coin}
c=\begin{pmatrix}
c_{\uparrow\uparrow}&c_{\uparrow\downarrow}\\
c_{\downarrow\uparrow}&c_{\downarrow\downarrow}
\end{pmatrix},
\end{equation}
where $\sum_{ij}|c_{ij}|^2=1$. A coin such that two players $A$ and $B$ have exactly the same probability of getting heads and tails can be defined as $|c_{\uparrow\downarrow}|=|c_{\downarrow\uparrow}|$. Using \eqref{eq:PlayerA} and \eqref{eq:PlayerB}, 
\begin{align}
\begin{aligned}
\text{Prob}_A(\uparrow)&=|c_{\uparrow\uparrow}|^2+|c_{\uparrow\downarrow}|^2=|c_{\uparrow\uparrow}|^2+|c_{\downarrow\uparrow}|^2 \\ &=\text{Prob}_B(\uparrow)\\
\text{Prob}_A(\downarrow)&=|c_{\downarrow\uparrow}|^2+|c_{\downarrow\downarrow}|^2=|c_{\uparrow\downarrow}|^2+|c_{\downarrow\downarrow}|^2 \\ &=\text{Prob}_B(\downarrow)
\end{aligned}
\end{align}

In addition to $|c_{\uparrow\downarrow}|=|c_{\downarrow\uparrow}|$, they can play a fair coin by adding  $|c_{\uparrow\uparrow}|=|c_{\downarrow\downarrow}|$: 
\begin{align}
\begin{aligned}
\text{Prob}_A(\uparrow)&=|c_{\uparrow\uparrow}|^2+|c_{\uparrow\downarrow}|^2=|c_{\downarrow\downarrow}|^2+|c_{\downarrow\uparrow}|^2 \\&=\text{Prob}_A(\downarrow)\\
\text{Prob}_B(\uparrow)&=|c_{\downarrow\uparrow}|^2+|c_{\uparrow\uparrow}|^2=|c_{\uparrow\downarrow}|^2+|c_{\downarrow\downarrow}|^2 \\&=\text{Prob}_B(\downarrow)
\end{aligned}
\end{align}

Hence the most general form of a fair coin~\eqref{eq:coin} is 
\begin{equation}
c=\begin{pmatrix}
\sqrt{\frac{a}{2}}e^{i\theta_{\uparrow\uparrow}}&\sqrt{\frac{1-a}{2}}e^{i\theta_{\uparrow\downarrow}}\\
\sqrt{\frac{1-a}{2}}e^{i\theta_{\downarrow\uparrow}}&\sqrt{\frac{a}{2}}e^{i\theta_{\downarrow\downarrow}}
\end{pmatrix},~0\le a\le1.
\end{equation}
The classical setting can be recovered by putting $\theta_{ij}=0$ for all $i,j$ and $a=\frac{1}{2}$. The non-trivial phases play important roles in an quantum extensive form game~\cite{ikeda2021quantum,2022arXiv220705435I}. 
Both parties should make as many coin states as possible before starting the game to check the probability distribution before playing. 

The problem \textcolor{red}{($\clubsuit$)} has already been solved for the following reasons: each player can confirm their opponent's result by measuring their own second qubit at the end of the game as shown in Fig.~\ref{fig:circuit}. In this game, only measuring one's own qubits is allowed, but one of the possible attacks from one player to the other are as follows: it is natural to ask what would happen if the second qubit were measured first. Suppose Player $B$ measures their second qubit before Player $A$ tosses the coin, thus finalizing Player $A$'s result. However, it turns out that this attack is meaningless, as we will see below. The probability that Player $B$ observes $\textcolor{red}{\ket{i}_B}$ is $\sum_{j}|c_{ij}|^2$, which is equal to the probability \eqref{eq:PlayerA} that Player $A$ tosses a coin themselves and observes $\textcolor{blue}{\ket{i}_A}$. Therefore, there is no incentive for Player $B$ to measure the second qubit first; the observation of Player $B$'s second qubit does not give either Player $B$ or $A$ any advantage or disadvantage.

The problem \textcolor{red}{($\spadesuit$)} no longer exists in our protocol. Due to entanglement in game state~\eqref{eq:initial}, as soon as the state $\textcolor{blue}{\ket{i}}$ of one player's coin is determined, the other player's qubit $\textcolor{red}{\ket{i}}$ used for confirmation is instantly determined. As is well known, this does not mean that information is being transmitted beyond the speed of light~\cite{PhysRev.47.777}.

\subsection{Solutions to Other Possible Attacks}
In this game, the quantum state given initially is never broken and is played to the end. Each player can only observe their own state, and the results of their observations do not affect others. In situations where only flipping (measuring) a coin is allowed for each player, there is no room for discussing the outcome of the game. Moreover, as both players are aware of the outcome on both sides, they have to accept the result. Thus, the game is perfectly fair and works well.

In the original setting of quantum coin flipping game, only measuring one's own qubits is allowed, however, as a general extension of the game, we can also consider the case where players can manipulate their own qubits. In this case it is possible for a player to claim that they have obtained a value different from their opponent's result $\textcolor{blue}{\ket{j}_B}$:
\begin{equation}
    \textcolor{red}{\ket{j}_A\ket{i}_B}\to \textcolor{red}{U_A\ket{j}_A\ket{i}_B}=\textcolor{red}{\ket{j'}_A\ket{i}_B}.
\end{equation}

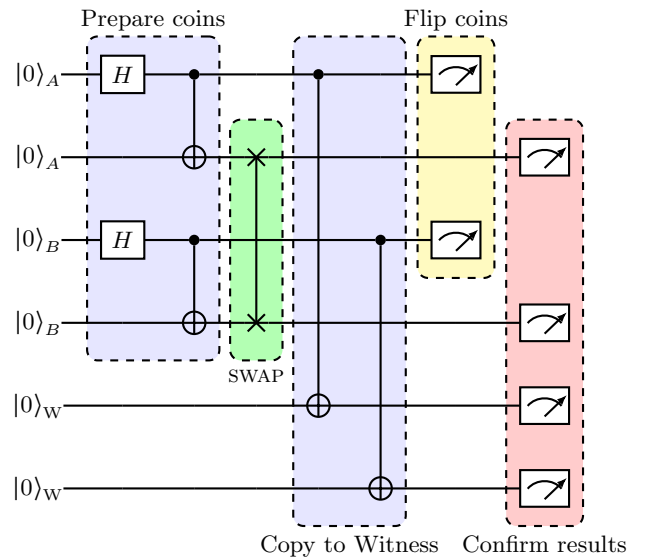
\begin{figure}[H]
    \centering
\begin{quantikz}
\ket{0}_A   &\gate{H}\gategroup[4,steps=2,style={dashed,rounded corners,fill=blue!10,inner xsep=2pt},background]{{Prepare coins}}& \ctrl{1} &\qw &\ctrl{4}\gategroup[6,steps=2,style={dashed,rounded corners,fill=blue!10,inner xsep=2pt},background, label style={label position=below,anchor=north,yshift=-0.2cm}]{{Copy to Witness}}&\qw& \meter{}\gategroup[3,steps=1,style={dashed,rounded corners,fill=yellow!30,inner xsep=2pt},background]{{ Flip coins}}\\
\ket{0}_A&\qw& \targ{} & \swap{2}\gategroup[3,steps=1,style={dashed,rounded corners,fill=green!30,inner xsep=2pt},background, label style={label position=below,anchor=north,yshift=-0.2cm}]{{\sc swap}}&\qw& \qw&\qw&\meter{}\gategroup[5,steps=1,style={dashed,rounded corners,fill=red!20,inner xsep=2pt},background, label style={label position=below,anchor=north,yshift=-0.2cm}]{{Confirm results}}\\
\ket{0}_B &\gate{H}& \ctrl{1} &\qw &\qw&\ctrl{3}&\meter{}\\
\ket{0}_B&\qw& \targ{} &\targX{} & \qw&\qw&\qw&\meter{}\\
\ket{0}_\text{W} &\qw&\qw&\qw& \targ{} &\qw&\qw &\meter{}\\
\ket{0}_\text{W}&\qw&\qw&\qw& \qw &\targ{} & \qw&\meter{}
\end{quantikz}
    \caption{Quantum Circuit for Quantum Coin Flipping Game with Witness. Here $\ket{0}_W$ is a qubit of Witness.}
    \label{fig:Witness}
\end{figure}

This issue is easily prevented by adding a third party (Witness) to the network. For this, we modify the initial state of coin \eqref{eq:initial} as follows
\begin{equation}
    \ket{\psi}_\text{Coin}=\sum_{ij} c_{ij} \overbrace{\textcolor{blue}{\ket{i}_A \ket{j}_B}}^{\text{Flipping}}\otimes\overbrace{\textcolor{red}{\ket{j}_A\ket{i}_B}}^{\text{Confirmation}}\otimes\overbrace{\ket{ij}_\text{Witness}}^{\text{Witness}}.
\end{equation}
Fig.~\ref{fig:Witness} shows a quantum circuit to play quantum coin flipping game with a witness for the case of constant $c_{ij}=\frac{1}{2}$. 

\begin{figure}[H]
    \centering
    \includegraphics[scale=0.35]{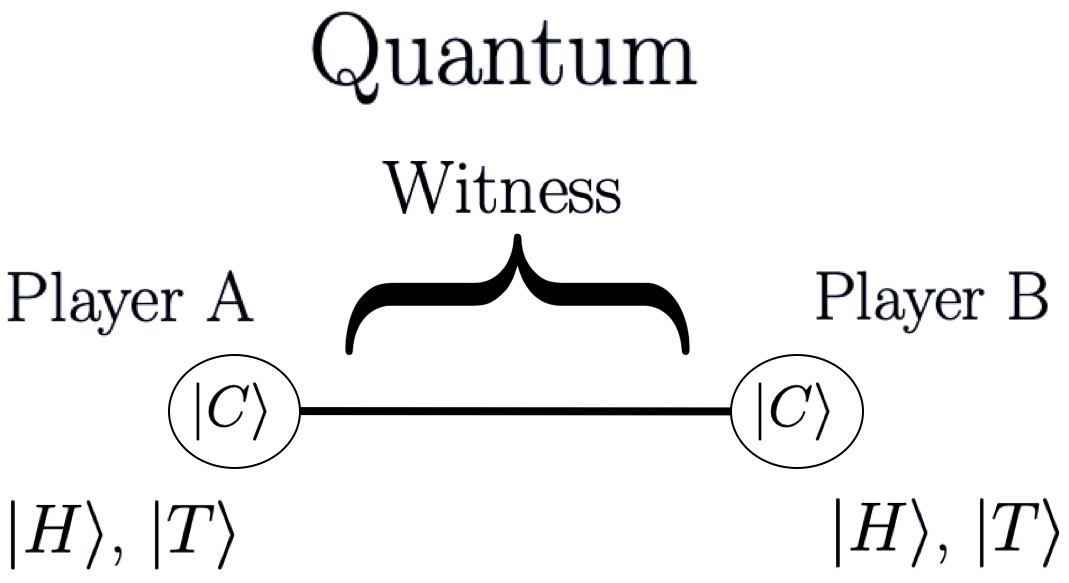}
    \caption{An extension of the quantum coin flipping game when manipulation to players' qubits are allowed can be represented by the figure above. Both players perform their respective quantum measurements on the quantum coin $\ket{C}$, and based on their measurements (either $\ket{H}$ or $\ket{T}$), the witness can verify the measurements. Subsequently, the players have no way of falsely communicating their results to each other.}
    \label{fig2}
\end{figure}
The first and second qubits of Witness corresponds to the states of Player $A$ and $B$, respectively. After all players flip their respective coins, the state of the game changes into 
\begin{equation}
\ket{\psi}_\text{Coin}\twoheadrightarrow\overbrace{\textcolor{red}{\ket{j}_A\ket{i}_B}}^{\text{Confirmation}}\otimes\overbrace{\ket{ij}_\text{Witness}}^{\text{Witness}}.
\end{equation}
Witness can confirm that Player $A$'s result is $i$ and Player $B$'s result is $j$, respectively. Even if Player A performs the same operation $U_A$ on their state as before, it does not change the record of the game held by Witness:
\begin{equation}
\textcolor{red}{\ket{j}_A\ket{i}_B}\otimes\ket{ij}_\text{Witness}\xrightarrow{U_A}\textcolor{red}{\ket{j'}_A\ket{i}_B}\otimes\ket{ij}_\text{Witness}.
\end{equation}
Ultimately, Player $A$ might try to change the state of Witness as well, but this is a completely different game, and is not in the scope of quantum coin flipping game. 

Thus, as long as the game is played via a proper quantum channel, there is no room for cheating on both sides (Condition \textcolor{blue}{(A)}). Each player is correctly aware of the other's results (Condition \textcolor{blue}{(B)}) and can agree on the winner of the game (Condition \textcolor{blue}{(C)}). Therefore the problems with the conventional quantum coin flipping game have been completely solved by redesigning the game and utilising a shared entangled state.

\section{$N$-party Quantum Coin Flipping Game}
\subsection{Motivation \& General Remark}
Here we extend our previous design of quantum flipping game for 2 persons. Before we present some explicit architectures, let us explain our motivations to consider $N$-person games. The problem in designing a two-person quantum coin flipping game was how to create a system in which two remote players could correctly share \textcolor{blue}{(B)} and agree on their true results \textcolor{blue}{(C)} without cheating \textcolor{blue}{(A)}. As we described in Introduction, this is a non-trivial task. 

In the case of $N$-player quantum flipping game, we present three solutions: central review, peer-to-peer review and hybrid peer-to-peer review. The first is the simplest method, but the most accurate and universal. As we did in the previous section, we invite an authorized third party (Witness) into the network. To eliminate unnecessary concerns, we assume that the state of Witness is not accessible from the outside. All participants must agree before the game begins that the result of the authorized Witness will be the final decision of the game.

\subsection{Design \& Solution 1: Central Review}
\begin{figure}[H]
    \centering
\begin{quantikz}
\ket{0}_1   &\gate{H}\gategroup[2,steps=1,style={dashed,rounded corners,fill=blue!10,inner xsep=2pt},background]{{Prepare Coins}}& \ctrl{2}\gategroup[4,steps=2,style={dashed,rounded corners,fill=blue!10,inner xsep=2pt},background, label style={label position=below,anchor=north,yshift=-0.2cm}]{{Copy to Witness}} &\qw & \meter{}\gategroup[2,steps=1,style={dashed,rounded corners,fill=yellow!30,inner xsep=2pt},background]{{ Flip coins}}\\
\ket{0}_2&\gate{H}& \qw & \ctrl{2}& \meter{}\\
\ket{0}_W &\qw& \targ{} &\qw &\qw&\meter{}\gategroup[2,steps=1,style={dashed,rounded corners,fill=red!20,inner xsep=2pt},background, label style={label position=below,anchor=north,yshift=-0.2cm}]{{Confirm results}}\\
\ket{0}_W&\qw& \qw &\targ{} & \qw&\meter{}
\end{quantikz}
    \caption{Central Review Quantum Circuit for Quantum Coin Flipping Game with $N$-person.}
    \label{fig:CentralReview}
\end{figure}
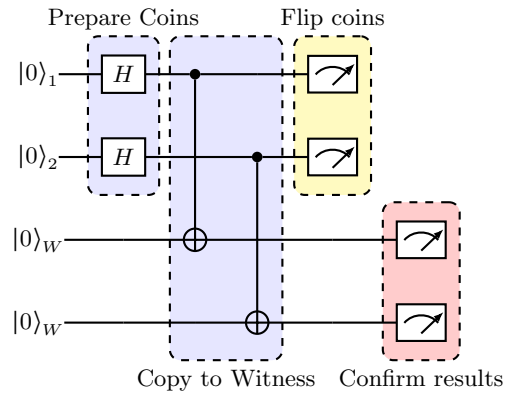

As an initial state of the game, all the players and Witness share the following entangled state
\begin{equation}
    \ket{\psi}_\text{coin}=\sum c_{i_1\cdots i_N}\overbrace{\textcolor{blue}{\ket{i_1\cdots i_N}}}^{\text{Flipping}}\otimes\overbrace{\ket{i_1\cdots i_N}_\text{Witness}}^{\text{Witness}},
\end{equation}
whereas before we use \textcolor{blue}{blue text} for coin qubits of the players. Fig.~\ref{fig:CentralReview} shows a quantum circuit to play quantum coin flipping game with a witness for the case of constant $c_{ij}=\frac{1}{2^{N/2}}$. 

Below is an example of a game in progress for the case $N=3$. When Player 1 first tosses the coin and observes $\textcolor{blue}{\ket{i_1}_1}$, the corresponding qubit $\ket{i_1}$ of Witness is decided uniquely. The same is true for Player 2 and 3 who get $\textcolor{blue}{\ket{i_2}_2}$ and $\textcolor{blue}{\ket{i_3}_3}$, respectively:
\begin{align}
\begin{aligned}
\ket{\psi}_\text{Coin}&\xrightarrow{\text{Player 1}}\sum_{j_2j_3}c_{i_1j_2j_3}\textcolor{blue}{\ket{j_2j_3}}\otimes\ket{i_1j_2j_3}\\
&\xrightarrow{\text{Player 2}}\sum_{j_3}c_{i_1i_2j_3}\textcolor{blue}{\ket{j_3}}\otimes\ket{i_1i_2j_3}\\
&\xrightarrow{\text{Player 3}}\ket{i_1i_2i_3}\\
\end{aligned}    
\end{align}
As in the $N=2$ case, Witness can check the results for all participants by measuring their own qubits one by one.
This can be further extended for the $N-$party scenario as follows. After $n(\le N)$ players flipped their coins and they observed $\{i_1,i_2,\cdots,i_n\}$ , the state of game changes into     
\begin{equation}
\begin{split}
\ket{\psi}_{\text{Coin}} \xrightarrow{{\text{$n$ players}}} &\sum_{
j_{n+1}\cdots j_{N}}c_{i_1\cdots i_nj_{n+1}\cdots j_N} \textcolor{blue}{\bigotimes_{m=n+1}^{N} {\ket{j_{m}}}}  \\ &\bigotimes_{m=1}^{n} \ket{i_{n}}\bigotimes_{m = n+1}^{N} \ket{j_{m}},
\end{split}
\label{N1}
\end{equation}
where $c_{i_1\cdots i_nj_{n+1}\cdots j_N}$ gives the coefficient for the rest of $N-n$ players after $n$ players flipped their coins. This is a generalisation of the previous examples, however this is only correct for $N-1$ players, therefore to complete the $N$-party game, the result for $n=N$ must be considered. This is given by
\begin{equation}
\ket{\psi}_{\text{Coin}} \xrightarrow{{\text{All players}}} \ket{i_1\cdots i_N}.
\label{N2}
\end{equation}
By combining equations (\ref{N1}) and (\ref{N2}), this gives the full solution for the central review protocol where $N$-players can confirm each others measurements through the witness. This is dependent on the assumption that in the \textbf{Decision making stage}, by prior agreement, the players agree on the Witness observation as the final outcome of the game.

The advantage of this method is the results are determined immediately after the game is over, since the Witness status is uniquely determined as soon as everyone flips a coin. Moreover this is the simplest system, requiring only $2N$~qubits and $2$ depth of gates to prepare the state before the game starts. However, if there is any doubt about the reliability of Witness or if there is an error in Witness' quantum measurement, an untrue result could be the final outcome of the game. 

\subsection{Design \& Solution 2: Peer-to-Peer Review}
\begin{figure*}
    \centering
\begin{quantikz}
\ket{0}_1   &\gate{H}\gategroup[9,steps=3,style={dashed,rounded corners,fill=blue!10,inner xsep=2pt},background]{{Prepare coins}}& \ctrl{1}&\ctrl{2} &\qw &\qw&\qw &\qw&\meter{}\gategroup[7,steps=1,style={dashed,rounded corners,fill=yellow!30,inner xsep=2pt},background]{{ Flip coins}}\\
\ket{0}_1&\qw& \targ{} & \qw&\swap{3}\gategroup[8,steps=4,style={dashed,rounded corners,fill=green!30,inner xsep=2pt},background, label style={label position=below,anchor=north,yshift=-0.2cm}]{{\sc swap}}&\qw& \qw&\qw&\qw&\meter{}\gategroup[8,steps=1,style={dashed,rounded corners,fill=red!20,inner xsep=2pt},background, label style={label position=below,anchor=north,yshift=-0.2cm}]{{Confirm results}}\\
\ket{0}_1&\qw& \qw & \targ{}&\qw&\qw&\qw&\swap{3}&\qw&\meter{} \\
\ket{0}_2 &\gate{H}& \ctrl{1} &\ctrl{2}&\qw &\qw&\qw&\qw&\meter{}\\
\ket{0}_2&\qw& \targ{} &\qw&\targX{} & \swap{3}&\qw&\qw&\qw&\meter{}\\
\ket{0}_2&\qw& \qw&\targ{}&\qw&\qw&\swap{3}&\targX{}&\qw&\meter{}\\
\ket{0}_3 &\gate{H}& \ctrl{1}&\ctrl{2} &\qw&\qw&\qw &\qw&\meter{}\\
\ket{0}_3&\qw& \targ{}&\qw &\qw & \targX{}&\qw&\qw&\qw&\meter{}\\
\ket{0}_3&\qw& \qw&\targ{}&\qw&\qw&\targX{}&\qw&\qw&\meter{}
\end{quantikz}
    \caption{Peer-to-Peer Quantum Circuit to prepare a state for $N$-Player Quantum Coin Flipping Game. For example, Player~1 reviews results $\{i_2,i_3\}$ of Player~2 and Player~3.  }
    \label{fig:p2p_circuit}
\end{figure*}
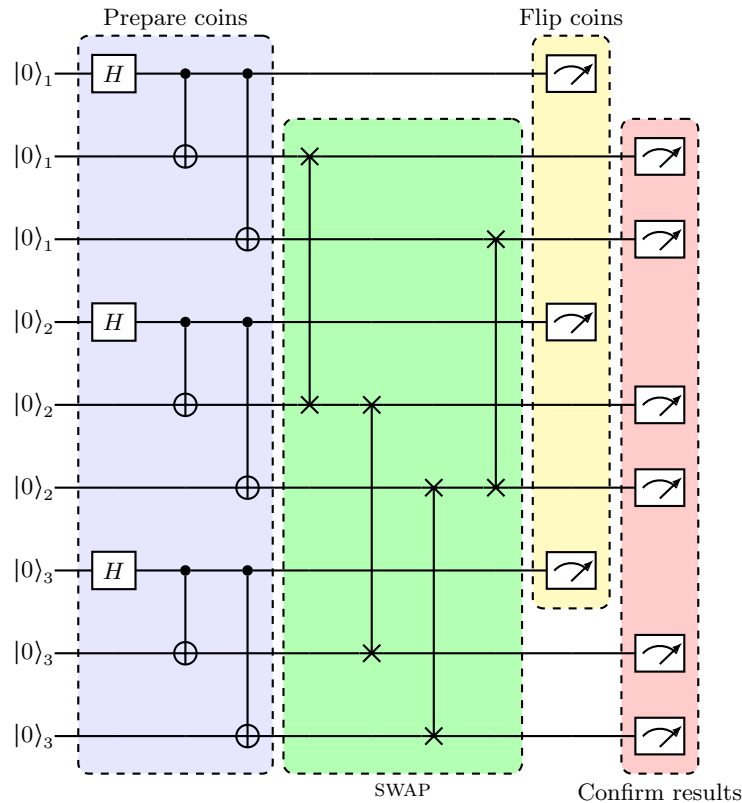

Here we provide a solution to $N$-player quantum coin flipping game without an authorized third party (Fig.~\ref{fig:p2p_circuit}). In this system, all participants review the results of other participants. Again we use \textcolor{blue}{blue text} for coin qubits and \textcolor{red}{red text} for confirmation qubits.

The simplest way to extend the state \eqref{eq:initial} used for the two-player game is to use 
\begin{equation}
    \ket{\psi}^{1}_\text{coin}=\sum c_{i_1\cdots i_N}\overbrace{\textcolor{blue}{\ket{i_1\cdots i_N}}}^{\text{Flipping}}\otimes\overbrace{\textcolor{red}{\ket{i_2i_3\cdots i_Ni_1}}}^{\text{Confirmation}}.
\end{equation}
For $i=1,\cdots,N-1$, Player $i$ confirms a result of Player~$(i+1)$ and Player $N$ confirms a result of Player~1. However, this leaves the verification of Player $i$'s results entirely up to Player $(i+1)$. This state can be prepared as illustrated in Fig.~\ref{fig:p2p_circuit}.

In order to achieve a peer-to-peer solution, we prepare the following state
\begin{align}
\begin{aligned}
\label{eq:p2p}
    \ket{\psi}_\text{coin}&=\sum c_{i_1\cdots i_N}\overbrace{\textcolor{blue}{\ket{i_1\cdots i_N}}}^{\text{Flipping}}\overbrace{\textcolor{red}{\bigotimes_{n=1}^N\ket{i_1\cdots i_{n-1}i_{n+1}\cdots i_N}}}^{\text{P2P Review}},
\end{aligned}
\end{align}
where at $n=1$ and $n=N$ the states should be understood as $\ket{i_2\cdots i_N}$ and $\ket{i_1\cdots i_{N-1}}$, respectively. This state can be prepared by operating SWAP operators between two different qubits for all combinations of Players.

Each player can know the results of everyone else by observing their own confirmation qubit $N-1$ times. For example, Player 1 obtains the set $\{i_2,i_3,\cdots,i_N\}$ of results of all others. The result of Player $i$ is reviewed by the other players. Let $i_n$ be a result of Player $n$ and let $i_{nj}$ the data of Player $n$ confirmed by Player $j$ observing their own qubit. A dataset $\{i_{n1},\cdots,i_{nN}\}$ for $N-1$ players regarding the results of Player $n$ is obtained. Let $r_n$ be the ratio that the result of Player $n$ agrees with the review results of other players
\begin{equation}
    r_{n}=\frac{\#\{j: i_n=i_{nj}\}}{N-1},
\end{equation}
where $\#$ denotes the number of elements in the set. Let $R_n$ be the ratio that the result of Player $n$ does not agree with the review results of other players
\begin{equation}
    R_{n}=\frac{\#\{j: i_n\neq i_{nj}\}}{N-1}.
\end{equation}
All participants decide before the game starts a constant criteria $r,R$, which do not depend on a particular $n$, for $r_n$ and $R_n$ to approve each player's result. 
For example, $i_n$ will be accepted if $r_n\ge r$, otherwise it will be rejected.

One advantage of employing this method is that the outcome of the game is not dependent on a particular third party. If it is desirable for players to decide the outcome in a democratic manner, this method can be used. One undesirable aspect of this method is that it takes a long time to get results, and multiple players can collude to get an incorrect result. Another problem is that in such cases, there is no place to complain about fraud. Moreover, as shown in Fig.~\ref{fig:p2p_circuit}, this system requires $N^2$~qubits and $(2N-1)$-depth of gates to prepare a state to play the game. Given that in the case of the Central Review system, the required gate depth is constant $(=2)$ regardless of the number of participants, and the number of required qubits is $2N$, the peer-to-peer review system is much more expensive to implement.

\subsection{Design \& Solution 3: Hybrid Peer-to-Peer Review}
Here we consider a hybrid peer-to-peer review system as a complementary mechanism to the central review system and the peer-to-peer review system. This is beneficial for networks that require a central server with peer-to-peer capabilities. If necessary, it is possible to operate only a central server or only a peer-to-peer network. In this system the players use states \eqref{eq:p2p} with a state of Witness: 
\begin{align}
\begin{aligned}
\ket{\psi}_\text{coin}=&\sum c_{i_1\cdots i_N}\overbrace{\textcolor{blue}{\ket{i_1\cdots i_N}}}^{\text{Flipping}}\\
&\underbrace{\textcolor{red}{\bigotimes_{n=1}^N\ket{i_1\cdots i_{n-1}i_{n+1}\cdots i_N}}}_{\text{P2P Review}}\underbrace{\ket{i_1\cdots i_N}}_{\text{Central Review}}
\end{aligned}
\end{align}
This system is easily implementable by combing quantum circuits shown in Fig.~\ref{fig:CentralReview} and Fig.~\ref{fig:p2p_circuit}. The $N=2$ case is illustrated in Fig.~\ref{fig:Witness}.

There are two main ways to build consensus:
\begin{enumerate}
    \item Players will primarily follow Witness's results, but will appeal using peer-to-peer review results.
    \item Players will primarily follow peer-to-peer review results, but will appeal using Witness's results. 
\end{enumerate}
Witness's states are determined as soon as all players have flipped their coins, but the result of the peer-to-peer review is not available until all players have completed all measurements. Players can choose the first method if efficiency is a priority, or the second method if democracy is a priority.

\section{Discussion \& Future Directions}
The work presented in this paper opens up a wide range of avenues to pursue in the future. To the best of the authors' knowledge, this is the first study of quantum coin flipping games motivated by mechanism design and incomplete contracts.
While there has been much technical and theoretical research on quantum cryptography, there has been little discussion on what kind of systems/software are user-friendly. However, in order to promote and develop quantum computers and quantum communications in general society, research from this perspective is essential. Subsequently, quantum game theory will become increasingly important.

So far we investigated the quantum coin flipping game with pure states, but it will be interesting to extend the game to mixed states. For example, in this paper we focused on using entanglement to prevent cheating in quantum games, however it would be interesting to see if quantum discord could be utilised \cite{L_Henderson_2001,PhysRevLett.88.017901}. It has already been shown that quantum discord could be measured in a bipartite system \cite{lowe}, therefore it opens up the possibility of using quantum discord for quantum advantage. Furthermore, it would be particularly interesting if we were able to develop a protocol which could verify each players measurements, without each player having to specifically reveal their measurement. This could be done using a quantum zero knowledge proof~\cite{qzk}. This would be in the form of peer-to peer-review, however the players would be allowed to keep their measurements secret. This could be a realistic scenario if the players measurements reveal sensitive information. From this perspective, creating a generic (hybrid) peer-to-peer quantum system is also an interesting open question. 

The game could also be developed into a repeated game where the players play multiple times~\cite{2020QuIP...19...25I,2021QuIP...20..387I,2022arXiv220705435I}. This type of game could be used to reveal the distribution of the shared state between the players. Such a scenario may occur if the players are unaware of what shared state they are performing their measurements on and they would like to deduce the distribution of the shared state.

\section*{Acknowledgement}
KI thanks Dmitri Kharzeev for useful discussion about quantum games. Work of KI was supported by the U.S. Department of Energy, Office of Science, National Quantum Information Science Research Centers, Co-design Center for Quantum Advantage (C2QA) under Contract No.DESC0012704. AL thanks Igor Yurkevich and David Lowe for insightful discussions about the implementation of quantum mechanics in game theory. 

\bibliographystyle{utphys}
\bibliography{ref}
\end{document}